\documentclass[prb,twocolumn,aps,superscriptaddress,showpacs,floatfix]{revtex4}
\usepackage{amsmath}
\usepackage{graphicx}

\begin{document}
\title{Coherent emission from disordered arrays of driven Josephson vortices}
\author{Fabio Marchesoni}
\affiliation{Dipartimento di Fisica, Universit\`a di Camerino,
I-62032 Camerino, Italy}
\affiliation{Advanced Science Institute, The Institute of Physical
and Chemical Research (RIKEN), Wako-shi, Saitama, 351-0198, Japan}
\author{Sergey Savel'ev}
\affiliation{Advanced Science Institute, The Institute of Physical
and Chemical Research (RIKEN), Wako-shi, Saitama, 351-0198, Japan}
\affiliation{Department of Physics, Loughborough University,
Loughborough LE11 3TU, United Kingdom}
\author{Masashi Tachiki}
\affiliation{National Institute for Materials Science, 1-2-1 Sengen,
Tsukuba 305-0047, Japan }
\author{Franco Nori}
\affiliation{Center for Theoretical Physics, Department of Physics,
University of Michigan, Ann Arbor, MI 48109-1040, USA}
\affiliation{Advanced Science Institute, The Institute of Physical
and Chemical Research (RIKEN), Wako-shi, Saitama, 351-0198, Japan}
\date{\today}
\begin{abstract}
We propose a mechanism of coherent emission from driven vortices in
stacked intrinsic Josephson junctions. In contrast to
super-radiance, which occurs only for \emph{highly ordered} vortex
lattices, we predict resonant radiation emission from \emph{weakly
correlated} vortex arrays. Our analytical results for the THz wave
intensity, resonance frequencies, and the dependence of THz emission
power on dissipation are in good agreement with the ones obtained by
recent simulations.
\end{abstract}
 \maketitle

\section{Introduction}

It has been experimentally observed \cite{ustinov1,ustinov2,welp}
and confirmed both analytically \cite{emit1,emit2} and numerically
\cite{tachiki,tachiki2} that moving Josephson vortices (JVs) emit
sub-THz electromagnetic radiation. Tera-Hertz radiation has
applications in physics, astronomy, chemistry, biology, and
medicine~\cite{tera-appl}. This motivates recent
proposals~\cite{review-JJ-JPW} for THz filters \cite{prl,gold},
detectors \cite{yam}, quantum devices \cite{quant,quant1}, and
emitters \cite{emit2,tachiki} based on highly anisotropic layered
superconductors (e.g., $\rm Bi_2Sr_2CaCu_2O_{8+\delta}$), that can
be modelled as coupled intrinsic Josephson junctions (IJJ).

The ultimate challenge in this field is to produce coherent THz
radiation. It is commonly believed that this goal can be achieved
by controlling super-radiance from highly ordered vortex lattices
\cite{machida,pohang}. A vortex lattice is deemed necessary
because the constructive interference of Josephson plasma waves
from individual JVs is strongly suppressed by small amounts of
disorder. Unfortunately, driven periodic lattices are often very
unstable (especially in the presence of impurities, defects, and
pinning centers), and moving JVs form either a mixture of
coexisting different lattices \cite{kosh} or even disordered
arrays \cite{tachiki}. Moreover, a broad radiation spectrum by
individual vortices results in a broad spectrum of the emitted
radiation (e.g., Ref. \onlinecite{emit2}), in contrast to a
desirable resonant IJJ, where coherent radiation is characterized
by sharp spectral lines.

In this context, recent interesting simulations by Tachiki and
coworkers \cite{tachiki} show that coherent radiation may be
generated by JVs moving as disordered arrays, instead of just
ordered ones. This raises the question as under what conditions
JVs in layered superconductors emit coherent radiation. Solving
this problem is crucial to the effective design of $\rm
Bi_2Sr_2CaCu_2O_{8+\delta}$--based THz emitters.

The experimental demonstration\cite{welp} of THz radiation in zero
magnetic field and various failed attempts at detecting THz
emission in the presence of magnetic fields cast serious doubts on
the initial idea that moving JVs can radiate in this frequency
domain. Indeed, the now prevailing interpretation is that JVs
ought to be considered as perturbing degrees of freedom, which
destroy the layer coherence and thus cause the suppression of THz
radiation. In this study, however, we reach the conclusion that,
under appropriate conditions, applied magnetic fields do help
amplify and tune THz emission.  This interesting result is also
consistent with the recent systematic studies in
Ref.~\onlinecite{rakhmanov_2009}.

Below, we show that the \emph{nonlocal} nature of JVs in layered
superconductors is responsible for a two-scale dynamics. A longer
scale, $\lambda_{EM}$, characterizes the inter-vortex magnetic
interaction; spatial dispersion, or disorder, of vortices up to such
a scale has \emph{no} appreciable impact on the radiation mechanism.
In other words, in contrast with super-radiance, which is suppressed
by vortex disorder, in our approach radiation coherence is preserved
even if the vortex distribution can become appreciably modulated by
the radiation itself for wavelengths shorter than $\lambda_{EM}$.
The shorter relevant length scale $\lambda_{G}$ determines the
cross-section of the nonlinear vortex core (where the linear
approximation $\sin\varphi\approx\varphi$ with the gauge invariant
phase difference across the junction having a vortex is not valid).
According to this picture, for $\lambda_{G} \ll \lambda_{EM}$ and
for a sufficiently high vortex density, radiation is emitted through
a linear mechanism, as from a JV lattice, whereas the
vortex-radiation coupling occurs mainly in the inner JV cores. Under
these conditions, the magnetic interaction among vortices is much
weaker than their interaction with the emitted radiation. Our
approach explains the spatial modulation of the JV density
numerically found in Ref.~\onlinecite{tachiki,tachiki2}. Our
analytical estimates, based on a one-dimensional sine-Gordon (sG)
model, prove to be in good agreement with their simulations and
explain their results.

Let us now summarize a central idea of our approach. Consider a
moving JV lattice emitting radiation. This radiation will bounce
back and forth the sample  edges, like in a laser cavity. This
radiation accumulates and creates a standing wave with a
wavelength about $\lambda_{EM}$. This standing wave modulates the
JV density which is now in resonance with the standing wave. This
positive feedback enhances the radiation of vortices. Namely, the
JV motion emits radiation, which is weaker at first. This
radiation bounced inside the sample (acting as a cavity) locks the
collective motion of the JVs. This collective motion produces
stronger emission. The JVs then interact more strongly with the
electromagnetic standing wave, compared with the now much weaker
vortex-vortex interaction. Thus, the triangular vortex lattice,
produced by the vortex-vortex interaction, is finally replaced by
a more disordered, but still modulated by the radiation, vortex
structure.

\section{Nonlocal sine-Gordon model}

Layered superconductors can be considered as stacks of strongly
interacting IJJs. As the superconducting layers are only a few
nanometers thick, i.e., the inter-layer distance $s$ is much
smaller than the magnetic field penetration depth $\lambda_{ab}$,
the currents flowing through different junctions are coupled. On
neglecting, for the time being, external drives and internal
dissipation, a system of stacked IJJs is well described by the
coupled sine-Gordon equations~\cite{sinG},
\begin{equation}\label{sG}
 \left(1-\frac{\lambda_{ab}^2}{s^2}\Delta^2_n\right)
\left(\frac{\varphi^{(n)}_{tt}}{\omega^2_p}+\sin\varphi^{(n)}\right)
-\lambda^2 \varphi^{(n)}_{xx}=0,
\end{equation}
where  $\varphi^{(n)}$ is the gauge invariant phase difference
across the $n$th junction. Here, $\omega_p$ is the Josephson plasma
frequency, $\lambda$ the London penetration depth
($\lambda/\lambda_{ab}=\gamma$) along the layers, and the operator
$\Delta^2_n$ is defined by $\Delta^2_n
f=f^{(n+1)}-2f^{(n)}-f^{(n-1)}$.

A full analysis of this set of equations is a complicated problem
which requires numerical simulation. However, if we restrict
ourselves to the case of moderate magnetic fields, when JV cores
do not overlap, we can reduce Eq.~(\ref{sG}) to an effective 1D
problem. Indeed, as was shown in \cite{photon-prb} (see Eqs. (21) and (22) and Fig. 2 there), the phase difference $\varphi$ decreases very fast away from the junction where a vortex located.
%
%
Thus, a reasonable strategy could consist in neglecting the nonlinear couplings between junctions at a distance of some $s$ from the vortex center; on solving the linearized equations (\ref{sG}) for such junctions, one would end
up with a few coupled nonlinear equations for a few junctions in
the vicinity of the vortex center. The case when the nonlinearity
was restricted to one junction only has been considered in
Ref.~\onlinecite{emit2}: The coupled junction system of
Eq.~(\ref{sG}) boils down to a 1D Josephson junction described by
a nonlocal sine-Gordon equation. We assume below that retaining
the nonlinear coupling between more junctions can lead to the same
nonlocal 1D sine-Gordon equation with additional noise-like weak
perturbations.

For simplicity, let us consider the pair of adjacent junctions $j$
and $(j+1)$ locating a moving JV. We then reduce the description
of the IJJ stack to a 1D problem by assuming the nonlinear
coupling to be important only for the paired junctions and
linearizing Eq.~(\ref{sG}) for all other junctions (i.e., for $n
\neq j,j+1$). It is interesting to note that the importance of the
interaction between two neighboring junctions is numerically well
established \cite{lin}. Moreover, Koshelev \cite{koshelev}
recently reduced the multi-junction system to two coupled
junctions, and this model reproduced the simulation data in
Ref.~\onlinecite{lin} and interpreted the experimental results of
Ref.~\onlinecite{welp}.

Following the approach in Refs.~\onlinecite{gurevich,emit2,barone},
the equation for the averaged phase difference across a junction
pair $\varphi=(\varphi^{(j+1)} +\varphi^{(j)})/2$ can be written as
\begin{equation}
\label{gurevich}
\frac{\varphi_{tt}}{\omega_p^2}+\sin\varphi=\frac{\gamma
s}{2\pi}\int dx'
K_0\left(\frac{|x-x'|}{\lambda}\right)\varphi_{xx}(x') +P[\psi]\;
\sin\varphi,
\end{equation}
where $K_0$ is the modified Bessel function and
$$P[\psi]=1-\cos\psi$$
with $\psi=(\varphi^{(j+1)} -\varphi^{(j)})/2$. The length
\begin{equation}
\lambda_G \equiv \frac{\gamma s}{2}=\frac{\lambda_{EM}^2}{\lambda}
\end{equation}
defines the size of the JV core. Again, the contribution of the
next-to-neighbor junctions to the dynamics of the tagged JV
weakens fast \cite{photon-prb} with their distance from the vortex center, thus, allowing all other nonlinear equations
(\ref{sG}) to be replaced by an effective nonlinear medium.

An additional equation for $\psi$ can be derived for a pair of JVs
in two adjacent junctions, so that the equations for $\varphi$ and
$\psi$ form a closed set~\cite{gurevich,emit2}. However, when
extending Eq.~(\ref{gurevich}) to describe the collective motion of
$N$ travelling JVs (randomly distributed along $N_l$ stacked IJJs of
length $L$), the phase $\varphi$ can be regarded as a mean-field
superposition of the $n=N/N_l$ individual JVs phases,
$\varphi^{(i)}_v$, contained in one layer, only.

In our one-IJJ description we assume that $\psi$ is relatively small; 
this may be the case, for instance, due to the random superposition of the vortex dynamics in different junctions. Anyway, the good agreement between the analytical results reported here and earlier numerical simulations, validates {\it a posteriori} our assumption.
Thus, the functional $P[\psi]$ can be modelled as a spatial
perturbation $\delta+\epsilon P(x)$, where the real function
$P(x)$ can be either periodic or random in $x$, depending on the
operating conditions.
The constant $\epsilon$ is a measure of the
strength of the perturbation, while the small offset $\delta$ can
be conveniently eliminated by rescaling the dimensional parameters
$\omega_p$ and $\lambda_G$, as appropriate. Here, we assimilate
such a perturbation as an effective quenched Gaussian disorder
along the IJJs; that is, $P(x)$ is modeled as a random,
delta-correlated function with
\begin{equation}
\label{disorder.P} \langle P(x)\rangle=0, ~~~~\langle
P(x)P(x')\rangle = 2\delta(x-x'),
\end{equation}
and $\langle \dots \rangle$ denoting the average over different
disorder realizations. The constant $\epsilon$ will be taken as a
perturbation parameter and only effects to leading order in
$\epsilon$ will be considered. Moreover, deviations from the
Gaussian statistics, implicit in the definition of $P[\psi]$, are
assumed to be negligible within this approximation.

\subsection{Josephson vortex array}

We now introduce dimensionless units by expressing $x$ and $t$ in
units of the characteristic length $\lambda_{EM}$ and the reciprocal
of the plasma frequency $\omega_p$, respectively; that is
$$ x \rightarrow \tilde x=x/\lambda_{EM}, ~~~ t \rightarrow \tilde
t=\omega_{p} t.$$ As a consequence, the system characteristic
lengths $\lambda$ and $\lambda_{G}$ get rescaled as follows: $$
\lambda\rightarrow \tilde \lambda = \lambda/\lambda_{EM}, ~~~
\lambda_G\rightarrow \tilde \lambda_G =
\lambda_{EM}/\lambda=1/\tilde \lambda,$$ and, of course,
$\lambda_{EM} \rightarrow \tilde \lambda_{EM} = 1.$ Correspondingly,
$\varphi(x,t)$ is given in units of the magnetic flux quantum
$\phi_0$ and all speeds in units of $\omega_p\lambda$.
Hereafter, for the sake of simplicity, we shall only use
dimensionless variables and, therefore, omit the ``tilde" notation
altogether.

The field $\varphi(x,t)$, corresponding to a dense distribution of
JVs traveling with speed $V \ll 1$, can be expanded as
\cite{zakarov,kivshar}
\begin{equation}
\label{array} \varphi(x,t)= p\;(x-Vt) - \kappa \sin [p\;(x-Vt)]+
\dots
\end{equation}
where $p=2\pi \rho$, and $\rho=n/L$ denotes the linear JV density
with number $n$ of vortices located along the length $L$. The linear
term in Eq.~(\ref{array}) corresponds to the phase difference of a
uniform vortex spatial distribution, whereas the periodic correction
accounts for a residual phase modulation on the lattice scale
$1/\rho$, with amplitude $\kappa$ to be determined
self-consistently.  In the expansion (\ref{array}) we assume high JV
densities, $p \gg 1$, and small amplitudes $\kappa$. On inserting
expansion (\ref{array}) for $\varphi(x,t)$, the nonlocal field
equation (\ref{gurevich}) can be approximated to an effective
sine-Gordon equation, where
\begin{equation}
\label{kernel}\frac{1}{\pi \lambda} \int K_0\left(
\frac{|x-x'|}{\lambda}\right )\varphi_{xx}(x')\;dx'\;
\rightarrow\; {c_p}^2 \,\varphi_{xx},
\end{equation}
$${c^2_p}=[1+(p\lambda)^2]^{-\frac{1}{2}},$$
and, consistently,
$$\kappa = (\gamma_V/c_p p)^2,$$
with
$$\gamma_V=(1-V^2/c^2_p)^{-\frac{1}{2}}.$$
In the regime considered in Refs.~\onlinecite{tachiki,tachiki2},
where $\lambda \gg 1$ and $\gamma_V \simeq 1$, the parameter $c_p$
can be further approximated to $(p\lambda)^{-\frac{1}{2}}$.

The validity condition for truncating the expansion (\ref{array})
to its first order, $\kappa \ll 1$, or equivalently $c_p p \gg 1$,
implies a direct core-core interaction; that is $1/\lambda \gg
1/p$. We recall that here $1/\lambda$ represents the size of a
vortex core in dimensionless units. Accordingly, for $\lambda \gg
1$, $c_p$ must be regarded as the maximum velocity of a vortex
array in a layered superconductor, to be compared with the maximum
dimensionless velocity $1/\lambda$ of a single vortex (i.e.,
$\omega_p\lambda_G$ in dimensional units \cite{emit2,gurevich}).

In the opposite limit, $p/\lambda \ll 1$, vortices only weakly
interact on the magnetic length scale $\lambda_{EM}$ (rescaled
here to 1); the limiting velocity $c_p$ grows larger than
$1/\lambda$ and the JV array becomes unstable.

\subsection{Radiation mechanism}

The emission of radiation by fast moving JVs also takes place on the
magnetic length scale $\lambda_{EM}$. The effective phase difference
$\varphi$ associated with an array of JVs moving along an IJJ, thus,
obeys the perturbed local sine-Gordon equation
\begin{equation}
\label{sg} \varphi_{tt}-{c_p}^2 \varphi_{xx} +\sin \varphi= -\beta
\varphi_t -f + \epsilon P(x)\sin \varphi.
\end{equation}
Note that, in leading order, $\varphi=p\;(x-Vt)$, as can be seen
from Eq. (\ref{array}). Here, for completeness, we have restored
the viscous term $-\beta \varphi_t$ and the current-induced drive
$f$, that allow us to control the net JV speed $V$ (see, e.g.,
Ref.~\onlinecite{barone}).

Like in the more conventional single sine-Gordon-soliton
perturbation schemes \cite{currie,willis}, we consider the Ansatz
\begin{equation}
\label{phonon} \varphi(x,t) \rightarrow \varphi(x,t) + \chi(x,t),
\end{equation}
which, inserted in Eq. (\ref{sg}), yields \cite{kivshar}
\begin{equation}
\label{phonon.eq} \chi_{tt}-{c_p}^2 \chi_{xx} +(\cos \varphi)\chi=
-\beta \chi_t +\epsilon P(x)\sin \varphi,
\end{equation}
where $\varphi=0$ is the ground state and only terms
$\mathcal{O}(\epsilon)$ have been kept. The
wavenumber $q$ and the angular frequency $\omega$ of the
unperturbed plasmon modes (i.e., for $\beta=\epsilon=0$) form a
continuum spectrum \cite{currie}, with
\begin{equation}
\omega^2=1+{c_p}^2 q^2.
\end{equation}
 However, as for the field (\ref{array})
with $p\gg 1$, the radiation-vortex coupling $(\cos \varphi)\chi$
becomes negligible, and the plasma wave dispersion relation can be
approximated to $\omega={c_p}|q|$. On introducing the spatial
Fourier components of $P(x)$ and $\chi(x,t)$, defined by
$$P(x)=
\frac{2}{\pi}\int_{-\infty}^{\infty} P(k)e^{ikx}dk,\ \
~~\chi(x,t)= \frac{2}{\pi}\int_{-\infty}^{\infty}
\chi_q(t)e^{iqx}dq\;,$$
Eq.~(\ref{phonon.eq}) can be rewritten as \cite{malomed}
\begin{equation}
\nonumber \frac{d}{dt}B(q)-\frac{\beta}{2}B(q) =
\frac{\epsilon}{2i} \left [e^{i(\omega -pV)t}P(q-p) -(p \to -p)
\right ],
\end{equation}
where
$$B(q)\equiv (\dot\chi_q-i|q|\chi_q)e^{i\omega t}$$
is directly related to the spectral density of the array emission
power,
$$W(q)=\frac{4}{\pi} \frac{d}{dt}|B(q)|^2,$$
 that is
\begin{equation}
\label{sep} W(q)= \frac{2\epsilon^2}{\pi} \left [
\frac{|P(q-p)|^2\;(\beta/2)}{(\beta/2)^2 +(\omega-pV)^2} +(p \to
-p) \right ].
\end{equation}
Here we use that $P(x)$ is a real function, so that
$P^*(k)=P(-k)$. The notation ``$(p\to-p)$'' denotes the symmetric
term obtained by replacing $p \to -p$ in the first term inside the
square brackets. This means that two waves propagate in opposite
directions with the same frequency $\omega$; for
$|P(q-p)|=|P(q+p)|$, they generate standing plasma oscillations,
like those reported in Ref. \onlinecite{tachiki,tachiki2}. To
simplify our notation, hereafter we restrict ourselves to JVs
driven in one assigned direction, say $V>0$.

The spectral emission power (s.e.p.) (\ref{sep}) is key to our
analysis of a resonant IJJ. The spectrum $W(q)$ can be easily
specialized for any choice of $P(x)$. In the case of quenched
Gaussian disorder, see Eq. (\ref{disorder.P}),
\begin{equation}
\langle |P(k)|^2 \rangle=\frac{1}{8},
\end{equation}
 so that on disorder-averaging Eq.~(\ref{sep})
 we obtain the IJJ spectral emission power per unit of
length
\begin{equation}
\label{w.disorder} w(\omega)= \frac{\epsilon^2}{4\pi} ~\left
[\frac{\beta/2}{(\beta/2)^2 +(\omega-pV)^2} + (\omega \to
-\omega)\right ].
\end{equation}
This spectrum holds for $\beta \ll \omega$ or, equivalently, for
$V \gg \beta/p$, and has a sharp resonance maximum
\begin{equation}
\label{max.disorder} w^{\mathrm{max}}= \frac{\epsilon^2}{2\pi \beta}
\end{equation}
for
\begin{equation}
\label{res.frequency} \omega_r = p\;V.
\end{equation}

\subsection{Vortex dynamics}

Subject to a drive $f$ produced by an externally-applied
electrical current, the vortices in an IJJ flow with an average
speed $V$ and, simultaneously, their cores interact with the
electromagnetic waves they radiate. For the relatively weak vortex
core repulsion, $1/\lambda \gtrsim 1/p$, simulated in
\onlinecite{tachiki,tachiki2}, we expect that the vortex array can
be modulated, both in space and time, by the resonant plasma
modes.

To express the average speed $V$ of a JV array with $p/\lambda \gg
1$ as a function of the drive $f$, from Eq. (\ref{sg}) we derive the
energy balance equation per unit of length of radiating IJJ
\cite{balance}
\begin{equation}
\label{balance} w(V)\;+\;\beta(pV)^2\;=\;pV f.
\end{equation}
Equation~(\ref{balance}) tells us that the rate at which the drive
pumps energy into the system (right-hand-side), must be
equilibrated by the radiative, $w(V)$, and the viscous loss,
$\beta(pV)^2$, of the soliton array $\varphi(x,t)$
(left-hand-side).

For $V \gg \beta/p$, the total emission power of the radiating
sine-Gordon solitons,
\begin{equation}
w(V)=\frac{\epsilon^2}{2},
\end{equation}
 is computed by integrating the spectral
emission power (\ref{sep}); solving the ensuing
Eq.~(\ref{balance}) with respect to $V$, we obtain
\begin{equation} \label{IV.nonlinear}
V(f)=\frac{f \pm (f^2-2\beta \epsilon^2)^\frac{1}{2}}{2\beta p},
\end{equation}
where only the rising branch with the $+$ sign is stable
\cite{malomed}. Therefore, the observable velocity-drive
characteristic $V(f)$ is expected to show a step at
\begin{equation} \label{threshold}
f_{\mathrm{th}}=(2\beta)^{\frac{1}{2}}\epsilon
\end{equation}
 and to grow linearly
with $f$ for $f\gg f_{\mathrm{th}}$, when the radiation loss
becomes negligible, namely
\begin{equation}
\label{IV} V=\frac{f}{\beta p}.
\end{equation}
Note that, at variance with an emitting JV lattice \cite{malomed},
no multiple hysteretic steps in the $V(f)$ are predicted. Indeed,
the condition $f>f_{\mathrm{th}}$ simply implies that the effective
phase $\varphi$ is not pinned by disorder \cite{lowdamping}; for
$f\lesssim f_{\mathrm{th}}$, instead, the JV array can move only by
creeping, namely, through the nucleation and the subsequent
migration of array defects \cite{nucleation,nori1}. Creeping is
likely responsible for the smooth low-current $J$-${\cal V}$
characteristics shown in Fig. 4 of Ref.~\onlinecite{tachiki}.
Moreover, in the linear regime (\ref{IV}) the wavelengths
$\lambda_r$ of the emitted radiation are expected to be much shorter
than the length $L$ of the IJJ (see below), so that corrections due
to the appropriate standing-wave periodic boundary conditions are of
the order of $\lambda_r/L$.

A vortex is sensitive to the radiation field only when the
wavelengths $\lambda_r$ excited in the IJJ are larger than its
size. For the parameter choice of
Refs.~\onlinecite{tachiki,tachiki2}, this can only occur on the JV
core scale $\lambda_G$, because $\lambda_G \lesssim \lambda_r$.

The interaction between the radiation standing wave, say
\begin{equation}
\label{core.1} \chi(x,t)=\chi_0\cos(qx)\cos(\omega t+\phi),
\end{equation}
and a single JV solution of the nonlocal sine-Gordon equation
(\ref{gurevich}),
\begin{equation}
\label{core.2} \varphi(x)=\pi+2\arctan(\lambda x)
\end{equation}
(both in dimensionless units) is well described by the
nonrelativistic quasi-particle approach of
Ref.~\onlinecite{balance}. The JV center of mass with coordinate
$X(t)$ is subject to an oscillating sinusoidal trap
\begin{equation}
\label{core.3} \ddot X = -\beta \dot X +\chi_0
\frac{q^2}{\lambda}\;e^{-q/\lambda}\cos(qX)\cos(\omega t +\phi),
\end{equation}
with an amplitude which is exponentially suppressed at short
wavelengths, i.e., for $q/\lambda \gg 1$. However, for
sufficiently large trap amplitudes, the vortices in each layer get
spatially distributed with wavevector $q$.

\section{Comparison with numerical results}

The results in Refs.~\onlinecite{tachiki,tachiki2} can be easily
analyzed within the above theoretical framework. To make contact
with their numerical data, one must express: all lengths in units
of $\lambda$, with $\lambda=200\;\mu$m; the velocities in units of
the light speed in the dielectric $c=c_0/\sqrt{\epsilon_c}$, where
$c_0$ is the speed of light {\it in vacuo}, and $\epsilon_c=10$ is
the simulated dielectric constant; the forces in units of $J/J_c$,
where $J_c$ is the critical JJ current and $J$ is the
superconducting current across the IJJs; and the angular
frequencies in units of the plasma gap frequency
$\nu_p=\omega_p/2\pi=c/\lambda=0.47\times 10^{12}\;$Hz. Moreover,
the actual layer JV density is $\rho=n/L \simeq 0.6\;\mu$m$^{-1}$
with $L=100\;\mu$m, the layer thickness is $s=15\;$\AA, and the
penetration length ratio $\gamma=500$. For this choice of
numerical parameters, the length scales we introduced in the
previous section read, respectively, $\lambda_G=0.38\;\mu$m,
$\lambda_{EM}=8.7\mu$m and $1/p=0.27\;\mu$m.

First, we note that the simulations of
Refs.~\onlinecite{tachiki,tachiki2} correspond to the physical
condition where $\lambda_G \lesssim \lambda_r$. As for the
resonant modes $\chi_0 \propto \epsilon$, see
Eq.~(\ref{max.disorder}), the amplitude of the driving force in
Eq.~(\ref{core.3}) turns out to scale like $\epsilon
(\lambda/\lambda_G)^{1/2}$, which is strong enough to drag a JV
against the disorder field (\ref{disorder.P}) and the array of
restoring forces. This explains the disordered spatial
distribution of the emitting JVs, which, far from forming any
ordered lattice, seem rather to get trapped by the plasma standing
waves. In spite of the coherent nature of the plasma radiation,
the vortex distributions in each IJJ can differ from one another
because of the intrinsic disorder brought about by the layer-layer
coupling.

\begin{figure}[btp]
\centering
\includegraphics[width=8.0cm]{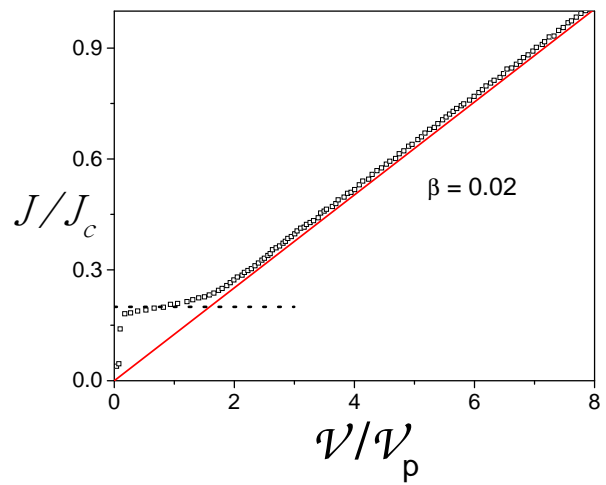}
\caption{(Color online) Current-voltage characteristics (in
dimensionless units) for a damping constant $\beta=0.02$. The
square symbols are the simulation data extracted from Fig.~4 of
Ref.~\onlinecite{tachiki}. The red straight line is our prediction
from Eq.~(\ref{IV.dim}) for the linear Ohmic branch. The
horizontal black dotted line is an estimate, from the numerics, of
the depinning threshold in Eq.~(\ref{threshold}).} \label{F1}
\end{figure}

In Fig. \ref{F1} we compare the current-voltage characteristics from
simulation, reported in Fig.~4 of Ref.~\onlinecite{tachiki}, with
the force-velocity ($f$-$V$) curve of Eq. (\ref{IV}). In the units
of Ref.~\onlinecite{tachiki}

\begin{equation}
\label{IV.dim} \frac{J}{J_c}=2\pi\beta\;\left (\rho
\lambda~\frac{V}{c}\right )=2\pi\beta\;\frac{\cal V}{{\cal V}_p},
\end{equation}

\noindent where ${\cal V}=\rho\lambda\; (V/c)$ is the flux-flow
voltage across a IJJ layer and ${\cal V}_p\equiv\nu_p\Phi_0$, with
$\Phi_0=h/2e$ denoting the flux quantum. The agreement is quite
good in the linear regime, whereas the depinning threshold
(\ref{threshold}) is clearly visible for $J/J_c\simeq 0.2$, which,
in our units, corresponds to setting $\epsilon =1$.

\begin{figure}[h]
\centering
\includegraphics[width=8.0cm]{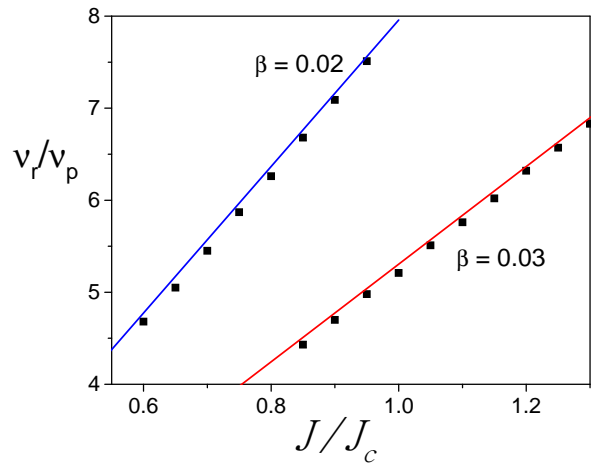}
\caption{(Color online) Resonance frequency $\nu_r$ versus current
intensity $J$ (both in dimensionless units) for two values of the
viscous constant $\beta$. The black dots are simulation data
extracted from Fig.~5 of Ref.~\onlinecite{tachiki}. The two
colored straight lines represent our theoretical predictions based
on Eq.~(\ref{JVspeed}), and using two different values of the
damping parameter $\beta$.} \label{F2}
\end{figure}

The resonant plasma radiation is investigated in
Refs.~\onlinecite{tachiki,tachiki2} on the linear branch of the
$J$-$\cal V$ characteristics. Three plots of the plasma standing
waves, two in Ref.~\onlinecite{tachiki} and one in
Ref.~\onlinecite{tachiki2}, are shown for different $J/J_c$; from
there we read out the corresponding resonance wavelengths
$\lambda_r$. Furthermore, the resonance frequencies
$\nu_r=2\pi\omega_r$ are either given explicitly in the text or
shown in the figure (see, e.g., Fig.~5 of
Ref.~\onlinecite{tachiki}): The product $\lambda_r \nu_r$ appears to
define a Swihart velocity, denoted here by $c_S$, independent of the simulation parameters
$J/J_c$ and $\beta$, that is
\begin{equation}
\frac{c_S}{c}\simeq 0.04.
\end{equation}
In view of our radiation mechanism
(\ref{phonon.eq}), the ratio $c_S/c$ can be identified with $c_p$ in Eq.(\ref{kernel});
accordingly, for $\lambda \gg 1$ and $\gamma_V \simeq 1$,
\begin{equation}
\frac{c_S}{c} \simeq \frac{1}{\sqrt{\lambda p}}
\end{equation}
is predicted to be of the order of
$0.036$, which is reasonably close to the result in Refs.~\onlinecite{tachiki,tachiki2},
given the accuracy of the data available.

The average JV speed in a resonant IJJ structure is proportional
to the resonance frequency, that is, from
Eq.~(\ref{res.frequency}),
\begin{equation}
\label{res.frequency.dim} V=\frac{\nu_r}{\rho}.
\end{equation}
This equation holds for {\it all} different choices of the
simulation parameters presented in
Refs.~\onlinecite{tachiki,tachiki2}. Note that the measured JV
speeds are relatively small, $V \ll c$, as assumed in our
nonrelativistic treatment of Eq.~(\ref{phonon.eq}), where
$\gamma_V\simeq 1$. Moreover, when combined with
Eq.~(\ref{IV.dim}), this equation yields the dependence of $\nu_r$
on the simulation control parameters $J/J_c$ and $\beta$. The
ensuing law
\begin{equation}
\label{JVspeed} \frac{\nu_r}{\nu_p}=\frac{1}{2\pi
\beta}\,\frac{J}{J_c}
\end{equation}
\\
closely matches all spectral resonance peaks reported in
Ref.~\onlinecite{tachiki}, as shown in Fig. \ref{F2}. Note that
combining Eqs. (\ref{IV.dim}) and (\ref{JVspeed}) yields the simple
$\beta$-independent relation
\begin{equation}
\frac{\nu_r}{\nu_p}=\frac{\cal V}{{\cal V}_p}.
\end{equation}
Finally, we notice from Eqs.~(\ref{max.disorder}) and (\ref{IV})
that $w^{\mathrm{max}}$ is proportional to $\epsilon^2/\beta$ and
$V$ is proportional to $f/\beta$; as a consequence, one would expect
that on decreasing $\beta$ the IJJ spectral emission band shifts to
lower $J/J_c$ while growing in intensity, both inversely
proportional to $\beta$. This is exactly the dependence displayed in
Fig.~6 of Ref.~\onlinecite{tachiki}.

\section{Conclusions}

We propose a new mechanism of coherent radiation from the moving
Josephson vortices in layered superconductors. We show, that due to
the two-scale structure of Josephson vortices, they radiate THz
radiation on a characteristic scale $\lambda_{EM}$, which is much
longer than the Josephson vortex core size $\lambda_G\sim \gamma s$.
Among all emitted waves, only standing modes in the sample (working
as a cavity) survive. These standing modes produce modulation of the
density of JVs. This, in turn, make vortices mainly radiate with
wavelengths corresponding to standing waves. Such positive feedback
can result in relatively strong radiation with well pronounced
maxima in the spectra. All our analytical estimates are in a good
agreement with numerical data~\cite{tachiki,tachiki2}.

The experimental demonstration\cite{welp} of THz radiation in zero
magnetic field and various failed attempts at detecting THz
emission in the presence of magnetic fields cast serious doubts on
the initial idea that moving JVs can radiate in this frequency
domain. Indeed, the now prevailing interpretation is that JVs
ought to be considered as perturbing degrees of freedom, which
destroy the layer coherence and thus cause the suppression of THz
radiation. In this study, however, we reach the conclusion that,
under appropriate conditions, applied magnetic fields do help
amplify and tune THz emission.  This interesting result is also
consistent with the recent systematic studies in
Ref.~\onlinecite{rakhmanov_2009}.

\section*{Acknowledgments}

FN acknowledges partial support from the National Security Agency
(NSA), Laboratory for Physical Sciences (LPS), Army Research
Office (ARO), National Science Foundation (NSF) grant No.
EIA-0130383. FN and SS acknowledge partial support from JSPS-RFBR
06-02-91200, and Core-to-Core (CTC) program supported by the Japan
Society for Promotion of Science (JSPS). S.S. acknowledges partial
support from the UK EPSRC via Nos. EP/D072581/1 and EP/F005482/1.

\end{document}